\documentclass[11pt]{article}

\usepackage[letterpaper,margin=1in]{geometry}
\usepackage{microtype}
\usepackage{amsmath,amssymb}
\usepackage{booktabs}
\usepackage{siunitx}
\usepackage{graphicx}
\usepackage{subcaption}
\usepackage{longtable}
\usepackage{array}
\usepackage{setspace}
\usepackage{enumitem}
\usepackage[colorlinks=true,allcolors=blue]{hyperref}
\usepackage{float}
\usepackage[english]{babel}
\usepackage{chngcntr}
\usepackage{tikz}
\usetikzlibrary{positioning,arrows.meta,shapes.geometric,calc}

\graphicspath{{./figures/}{./si_figures/}}
\setlist[itemize]{leftmargin=*,itemsep=0.25em,topsep=0.25em}
\setlist[enumerate]{leftmargin=*,itemsep=0.25em,topsep=0.25em}
\newcommand{\secref}[1]{\hyperref[#1]{Section~\ref*{#1}}}
\newcommand{\subsecref}[1]{\hyperref[#1]{Section~\ref*{#1}}}
\newcommand{\figref}[1]{\hyperref[#1]{Figure~\ref*{#1}}}
\newcommand{\tabref}[1]{\hyperref[#1]{Table~\ref*{#1}}}
\title{How Eviction Court Governs: A Statistical Analysis of Bargaining, Templates, and Debt in Philadelphia}
\author{
  Marios Papamichalis\thanks{Human Nature Lab, Yale University, New Haven, CT 06511, \texttt{marios.papamichalis@yale.edu}}
  \and
  Regina Ruane\thanks{Department of Statistics and Data Science, The Wharton School, University of Pennsylvania, 3733 Spruce Street, Philadelphia, PA 19104-6340, \texttt{ruanej@wharton.upenn.edu}}
}
\date{}
\begin{document}
\maketitle

\begin{abstract}
We analyze downstream courtroom governance in Philadelphia eviction cases using 755{,}004 Municipal Court landlord--tenant records filed from 1969 through 2022. Post-filing case processing is organized by repeated courtroom relationships, judge and tenant-attorney regimes, reusable agreement templates, and repeated team--property units. Among both-represented, both-attorney-named cases, 58.2\% involve a plaintiff-side and tenant-side attorney pair that had appeared against one another in the prior year, and greater prior pair exposure predicts lower default, higher judgment-by-agreement, and higher served-writ rates. Judge-linked cases display statistically distinct baseline outcome, continuance, fee, and award regimes; tenant-attorney identity explains meaningful variance in both case outcomes and agreement terms. Settlement text is highly standardized: reusable templates explain strictness, waiver, lockout-trigger, payment-plan, deadline, and time-is-essence language far more strongly than raw attorney identity. Monetary burden concentrates in repeated plaintiff--attorney--property units. Assignment-cell support and balance audits indicate that judge-linked evidence reflects institutional heterogeneity rather than a clean judge lottery, and judge--triad interactions are not estimable in this docket. Eviction court emerges as a repeated institutional field that organizes bargaining, text, debt, and enforcement after cases enter the courtroom pipeline.
\end{abstract}

\section*{Significance Statement}
Eviction research typically anchors on two endpoints: filings and physical removals. We measure the courtroom process that connects them. Represented cases repeatedly involve the same lawyer pairs; judges and tenant attorneys operate under distinct courtroom regimes; written agreements are governed mainly by reusable templates; and debt production concentrates in repeated team--property units. Assignment-cell support and balance audits show that the Philadelphia docket does not sustain a broad clean judge-lottery design, and judge--triad cells are too sparse for stable interaction estimates. Separating institutional structure from unsupported causal claims yields a quantitative account of courtroom governance with an explicit causal boundary.

\section*{Keywords}
Administrative data; court records; high-dimensional fixed effects; variance decomposition; difference-in-differences; eviction; statistical applications.

\section{Introduction}\label{sec:introduction}

Eviction is typically measured at two endpoints: the filing of a case and the physical removal of a tenant. We study the institutional process between them. Once filed, a case moves through a courtroom system of lawyers, judges, settlement agreements, templates, fees, awards, writs, and enforcement events. Focusing on Philadelphia, we ask four questions. First, do the same plaintiff-side and tenant-side attorneys repeatedly face each other, and do those repeated encounters correspond to different case outcomes? Second, do judges and tenant attorneys vary in the conditions under which cases are resolved? Third, are settlement agreements drafted case by case, or built from reusable templates? Fourth, is monetary burden driven by individual actors, or by repeated plaintiff--attorney--property units?

Courtroom processing is both relational and standardized. Repeated cross-side attorney pairings are common and predict fewer defaults, more judgments by agreement, and more served writs of possession. Judges and tenant attorneys are systematically associated with distinct outcome, continuance, fee, award, and agreement patterns. Settlement language is organized mainly through reusable templates rather than free-form drafting. Fee burden and award-over-sought peak where repeated legal teams and repeated properties coincide, particularly in plaintiff--attorney--property triads.

Our approach is deliberately statistical and institutional rather than experimental. We estimate repeated relationships, actor-level regimes, template-level variance, and support conditions in a large administrative panel. Pair-level findings characterize repeated courtroom relationships, not a randomized effect of attorney-pair familiarity. Judge-level findings document sorting and baseline differences across judge-linked cases, not a clean judge lottery. Template-level findings describe how settlement text is organized, not the causal effect of individual clauses. Debt-level findings locate where monetary burden concentrates within the court process. Assignment-cell balance checks, support screens, and a Callaway--Sant'Anna recurrence analysis delimit which claims can and cannot be interpreted causally.

\secref{sec:literature} positions the analysis within the eviction, courts, representation, document, and debt literatures; \secref{sec:data} defines samples, measures, and designs; \secref{sec:results} presents the attorney-pair, judge, template, and debt findings; and \secref{sec:discussion} discusses implications and limitations.

\section{Literature}\label{sec:literature}

Eviction scholarship has shown that court involvement is a major institutional pathway through which poverty, housing instability, discrimination, and health consequences are reproduced \cite{hartman2003evictions,desmond2012eviction,desmond2015eviction,desmond2017evicted,desmond2017gets,desmond2015forced,greenberg2016discrimination}. Administrative-record research then clarified both the scale of formal eviction and the difficulty of comparing legal stages across jurisdictions \cite{gromis2022estimating,graetz2023comprehensive,collinson2024eviction,nelson2021evictions,porton2021inaccuracies}. Informal displacement, pandemic moratoria, and emergency legal interventions show why filings, judgments, writs, agreements, and removals should be analyzed as distinct procedural objects rather than collapsed into a single eviction measure \cite{zainulbhai2022informal,hepburn2023protecting,benfer2023covid,summers2025pathways}.

The courtroom focus draws on socio-legal theories of dispute transformation, settlement, judicial management, and repeat-player advantage. Court disputes are filtered through legal categories and negotiation routines before they appear as outcomes, and settlement can be shaped by institutional rules even when it appears private \cite{felstiner2017emergence,mnookin1978bargaining,priest1984selection,fiss1983against,resnik1982managerial,hardaway2026courthouse}. Repeat-player theory predicts that actors who appear frequently can build practical advantages through specialization and institutional familiarity \cite{galanter1974haves}. Housing-court scholarship documents precisely the setting in which such advantages are likely to matter: high-volume courts, limited adjudicative attention, extensive default, and asymmetric access to lawyers \cite{shanahan2022judges,bezdek1991silence,engler2010connecting,sabbeth2022eviction,sudeall2021praxis}. Evidence on representation further shows that lawyers can alter tenant outcomes, but not through a single uniform mechanism across all courts and service models, and procedural burdens such as travel can independently alter default and participation \cite{seron2001impact,ellen2021lawyers,cassidy2023effects,summers2022eviction,greiner2012limits}.

A separate legal-profession literature helps explain why the analysis treats both plaintiff-side and tenant-side attorneys as organized subfields. Professional legal work is segmented across client worlds, and eviction practice may be routinized by specialized plaintiff counsel, high-volume filing, and repeated courthouse relationships \cite{wilf2021assembly,aizman2025shadow,heinz1982chicago,heinz2005urban}. More general accounts of landlord--tenant law, legal expertise, and the coded architecture of law treat legal actors and forms as infrastructure: they do not merely apply rules, but help build the pathways through which rights, risks, and obligations are made durable \cite{rabin1983revolution,sandefur2015elements,pistor2019code,shanahan2022institutional}. Related work using the same Philadelphia eviction docket has examined plaintiff-side filing concentration, repeated filing addresses, and tenant recurrence \cite{papamichalis2026legal}, as well as the relationship between high-volume plaintiff-side attorneys and the rise of single-Appearance eviction cases in Philadelphia \cite{papamichalis2026high}. We build on that filing-side work by examining the downstream courtroom layer: repeated attorney-pair encounters, judge sorting, tenant-attorney heterogeneity, settlement templates, and monetary awards.

The bargaining and template modules are grounded in scholarship on legal documents, boilerplate, and court-produced settlement. Pleadings and notices do more than record administrative information; they can structure claims, defaults, and bargaining leverage \cite{bernal2020pleadings,radin2013boilerplate}. Work on eviction records, tenant screening, and court-record consequences shows that the written residue of an eviction case can matter even when physical removal is not observed \cite{kleysteuber2006tenant,reosti2020we,eisenberg2024record,brantley2025record,estes2025justice}. Related work on settlement terms and civil-court design highlights that written agreements may govern future obligations, defaults, and enforcement possibilities long after the hearing itself \cite{summers2024evicted,humphries2019does,summers2026settlements}. We distinguish attorney identity from template identity to ask whether agreement language reflects individualized bargaining or standardized form governance.

The monetary-obligations module connects courtroom processing to research on serial filing, legal costs, racialized property markets, and civil fees. Prior work shows that landlords may use eviction filings as part of rent-collection and tenancy-management strategies, and that repeated filing is concentrated across properties, neighborhoods, and ownership structures \cite{garboden2019serial,leung2021serial,immergluck2020evictions,hepburn2020racial,rutan2021concentrated}. Monetary sanctions, filing fees, legal costs, and court-recorded monetary obligations can shift financial burden to defendants and deepen inequality, especially in racialized housing markets \cite{gomory2023racially,ajayi2026landlord,brito2022racial,harris2010drawing,holland2011one,leibowitz2010repairing,summers2023civil}. The courtroom-process question therefore extends beyond who files: how do repeated lawyers, judges, templates, and team--property units organize the monetary layer that accompanies possession claims?

\section{Data, measurement windows, and empirical strategy}\label{sec:data}

\subsection{Data}\label{sec:data_data}

We analyze 755{,}004 Philadelphia Municipal Court landlord--tenant case records filed from 1969 through 2022. The source extract contains 798{,}726 rows; restricting to the 1969--2022 filing window drops 43{,}722 records and yields the 755{,}004-row analytic frame. The residential universe contains 747{,}125 cases. We restrict the analysis to the courtroom layer of that record: represented attorney pairings, judge-linked cases, tenant attorneys, judgment-by-agreement (JBA) text, templates, fee components, awards, writs, and service of the alias writ of possession (the procedural instrument under which the sheriff is authorized to execute physical removal in Philadelphia). The courtroom layer is the object of analysis; filing-side infrastructure is addressed elsewhere \cite{papamichalis2026legal}.

Courtroom samples are drawn from the full docket and overlap; each module uses the subset for which the required identifiers are observed. Repeated attorney-pair analyses use the both-represented and both-attorney-named sample ($n = 26{,}259$). Judge sorting and judge-regime analyses use the judge-linked residential sample ($n = 54{,}273$). Agreement-text analyses use 220{,}622 non-empty JBA texts, with 84{,}099 attorney-identified texts entering the stricter attorney and template tests. Pair-identified JBA text is much thinner and is used only diagnostically (121 texts and 106 pairs in the paper-strict sample). Fee and award analyses use the modern fee universe, where fee components are substantively informative.

\subsection{Measurement windows}\label{sec:data_windows}

\begin{table}[H]
\centering
\caption{Measurement windows used in the courtroom-governance study.}
\label{tab:windows}
\small
\begin{tabular}{p{0.34\linewidth}p{0.22\linewidth}p{0.34\linewidth}}
\toprule
Module & Primary window / universe & Rationale \\
\midrule
Cross-side attorney pairs & Both-represented and both-attorney-named cases; stable window begins 1999 & Requires informative attorney names on both sides. \\
Judge sorting and judge regimes & Judge-linked residential sample & Requires observed judge identifiers and case outcomes. \\
Tenant-attorney heterogeneity & Represented tenant-side cases with usable attorney identifiers & Measures tenant-side variation in outcomes and agreement terms. \\
Descriptive JBA text & Non-empty agreement texts, observed at scale from 1976 & Agreement language exists for a large but selected subset of JBA cases. \\
Attorney- and template-identified JBA text & Stable window begins 1988 & Requires usable attorney identifiers and repeatable normalized text forms. \\
Fee share and award-over-sought & Modern fee universe; stable window begins 2005 & Fee components are substantively informative only in the modern period. \\
\bottomrule
\end{tabular}
\end{table}

\subsection{Core measures}\label{sec:data_measures}

For represented cases, we construct a plaintiff-attorney by tenant-attorney pair identifier and count prior same-pair encounters in the preceding 12 months. We summarize judge sorting through judge-by-actor contingency tables and chi-square tests, and summarize judge, tenant-attorney, pair, template, address, plaintiff--attorney pair, and triad heterogeneity with residualized pooled ANOVA designs and \(\eta^2\) effect sizes.

For JBA text, we concatenate and normalize the agreement and order fields. Clause indicators identify deadlines, move-out language, payment-plan provisions, waiver clauses, and lockout triggers. The primary strictness index is the additive sum of these binary indicators (range 0--5, mean 0.742); the additive and first-principal-component scores correlate at $r = 0.722$ across the JBA-text universe (SI Table~\ref{tab:jba_features_appendix}). Collapsing normalized agreement texts into template IDs separates reusable forms from generic attorney style. Because pair-identified JBA text is sparse, we evaluate pair-level text claims using support counts and holdout performance.

For debt, fee share is defined for positive-award cases as
\[
\mathrm{FeeShare}_i =
\frac{\mathrm{AttorneyFees}_i+\mathrm{Costs}_i+\mathrm{OtherFees}_i}
{\mathrm{TotalAward}_i}.
\]
Award-over-sought is the ratio of total award to amount sought when both quantities are positive. Repeated team--property structure is measured with plaintiff--attorney pairs, addresses, and plaintiff--attorney--property triads.

\subsection{Empirical strategy}\label{sec:data_strategy}

Our empirical strategy is descriptive and institutional. We estimate binary outcomes with linear probability models and continuous outcomes with OLS, evaluate sorting with chi-square tests, and summarize actor-level or unit-level heterogeneity with residualized pooled ANOVA and \(\eta^2\) effect sizes. We treat overfitting as a first-order concern in sparse pair-level text data and admit pair-specific bargaining-text claims only when both support counts and holdout performance are adequate.

Before interpreting any judge effect, we assess in the judge/courtroom appendix whether the data can support a quasi-random-assignment design. We screen candidate assignment cells for both support and covariate balance. Broad cells retain more judge-linked cases but fail balance; narrower cells improve comparability only by sharply reducing support. We therefore treat judge-linked estimates as courtroom-regime evidence unless a cell simultaneously satisfies support, balance, and institutional-assignment criteria.

The only staggered-treatment design we report is a supplemental repeat-triad recurrence diagnostic. Let \(G_i\) denote the quarter in which plaintiff--attorney--property triad \(i\) reaches its second observed filing. For triad-quarter outcomes \(Y_{it}\), the appendix estimates group-time average treatment effects
\[
ATT(g,t)=E\{Y_{it}(g)-Y_{it}(\infty)\mid G_i=g\}
\]
using Callaway--Sant'Anna difference-in-differences (CSDID) with not-yet-treated and never-recurrent triads as controls \cite{callaway2021difference}. We use quarterly bins, \(\log(1+\text{count})\) outcomes, an event window from four quarters before to eight quarters after recurrence, and a doubly robust estimator with pointwise event-time inference \cite{sant2020doubly}. The recurrence panel contains 1{,}737 treated triads, 1{,}500 sampled never-recurrent controls, and 307{,}515 triad-quarter observations. Because recurrence is not randomly assigned, we report these estimates as supplementary diagnostics and interpret them only when the pretrend screen is not rejected.

\section{Results}\label{sec:results}

\begin{figure}[H]
\centering
\definecolor{govblue}{RGB}{52,88,138}
\definecolor{govgray}{RGB}{120,124,132}
\resizebox{\linewidth}{!}{%
\begin{tikzpicture}[
  font=\small,
  >={Latex[length=2.4mm,width=2.0mm]},
  endpoint/.style={
    draw=govgray, fill=govgray!8, very thick,
    rounded corners=2.5pt,
    minimum width=2.0cm, minimum height=5.2cm,
    align=center, inner sep=5pt
  },
  pillar/.style={
    draw=govblue, fill=govblue!7, very thick,
    rounded corners=2.5pt,
    minimum width=3.0cm, minimum height=5.2cm,
    inner sep=5pt
  },
  ptitle/.style={anchor=north, font=\bfseries\small, align=center},
  pstat/.style={anchor=south, font=\footnotesize, align=center, text width=2.7cm},
  flowarrow/.style={->, line width=0.7pt, govgray!85},
  node distance=4mm and 4mm
]

\node[endpoint] (filing) {%
  \textbf{Case filing}\\[8pt]
  \footnotesize 755{,}004\\ cases\\[3pt]
  \footnotesize Philadelphia\\ Mun.\ Court\\[3pt]
  \footnotesize 1969--2022};

\node[pillar, right=7mm of filing] (p1) {};
\node[ptitle] at (p1.north) [yshift=-5pt] {Repeated\\ lawyer pairs};
\begin{scope}[shift={($(p1.north) + (0,-1.85cm)$)}]
  \node[circle, draw=govblue, fill=govblue!25, minimum size=14pt, inner sep=0pt] (i1a) at (-0.5,0) {};
  \node[circle, draw=govblue, fill=govblue!25, minimum size=14pt, inner sep=0pt] (i1b) at (0.5,0) {};
  \draw[govblue, line width=1.4pt] (i1a) -- (i1b);
\end{scope}
\node[pstat] at (p1.south) [yshift=5pt] {%
  \textbf{58\%} of cases\\
  reuse a lawyer\\
  pair from the\\
  prior year};

\node[pillar, right=of p1] (p2) {};
\node[ptitle] at (p2.north) [yshift=-5pt] {Judges \&\\ tenant attorneys};
\begin{scope}[shift={($(p2.north) + (0,-1.85cm)$)}]
  \node[circle, draw=govblue, fill=govblue!25, minimum size=12pt, inner sep=0pt] at (0,0.20) {};
  \draw[govblue, line width=1.4pt, fill=govblue!15] (-0.65,0.02) rectangle (0.65,-0.08);
  \draw[govblue, line width=1.0pt] (-0.52,-0.08) -- (-0.52,-0.32);
  \draw[govblue, line width=1.0pt] (0.52,-0.08) -- (0.52,-0.32);
\end{scope}
\node[pstat] at (p2.south) [yshift=5pt] {%
  Judges receive\\
  attorneys non-\\
  randomly; regimes\\
  differ};

\node[pillar, right=of p2] (p3) {};
\node[ptitle] at (p3.north) [yshift=-5pt] {Settlement\\ templates};
\begin{scope}[shift={($(p3.north) + (0,-1.85cm)$)}]
  \draw[govblue, line width=1.0pt, fill=govblue!10] (-0.40, 0.20) rectangle (0.18,-0.16);
  \draw[govblue, line width=1.0pt, fill=govblue!18] (-0.28, 0.08) rectangle (0.30,-0.28);
  \draw[govblue, line width=1.0pt, fill=govblue!26] (-0.16,-0.04) rectangle (0.42,-0.40);
\end{scope}
\node[pstat] at (p3.south) [yshift=5pt] {%
  Reusable forms\\
  shape agreement\\
  language more\\
  than lawyer style};

\node[pillar, right=of p3] (p4) {};
\node[ptitle] at (p4.north) [yshift=-5pt] {Plaintiff--attorney\\ --property triads};
\begin{scope}[shift={($(p4.north) + (0,-1.85cm)$)}]
  \node[circle, draw=govblue, fill=govblue!25, minimum size=11pt, inner sep=0pt] (t1) at (0,0.20) {};
  \node[circle, draw=govblue, fill=govblue!25, minimum size=11pt, inner sep=0pt] (t2) at (-0.50,-0.32) {};
  \node[circle, draw=govblue, fill=govblue!25, minimum size=11pt, inner sep=0pt] (t3) at (0.50,-0.32) {};
  \draw[govblue, line width=1.4pt] (t1) -- (t2) -- (t3) -- cycle;
\end{scope}
\node[pstat] at (p4.south) [yshift=5pt] {%
  52{,}443 repeat\\
  team--property\\
  units concentrate\\
  fees and awards};

\node[endpoint, right=7mm of p4] (out) {%
  \textbf{Case outcomes}\\[8pt]
  \footnotesize Default\\[3pt]
  \footnotesize Judgment by\\ agreement\\[3pt]
  \footnotesize Served writ\\[3pt]
  \footnotesize Fees / awards};

\draw[flowarrow] (filing.east) -- (p1.west);
\draw[flowarrow] (p1.east) -- (p2.west);
\draw[flowarrow] (p2.east) -- (p3.west);
\draw[flowarrow] (p3.east) -- (p4.west);
\draw[flowarrow] (p4.east) -- (out.west);

\end{tikzpicture}%
}
\caption{Four-pillar architecture of post-filing courtroom governance in the Philadelphia eviction docket. Between case filing and case outcomes, four institutional mechanisms structure case flow: repeated cross-side attorney pairings (\subsecref{sec:results_pairs}); judge and tenant-attorney regimes (\subsecref{sec:results_judges}); standardized settlement templates (\subsecref{sec:results_bargaining}); and repeated plaintiff--attorney--property triads (\subsecref{sec:results_debt}). The summary under each pillar names the headline finding from the corresponding subsection; full estimates appear in Tables~\ref{tab:pair_field}, \ref{tab:judge_tenant}, \ref{tab:bargaining}, and~\ref{tab:debt}.}
\label{fig:framework}
\end{figure}
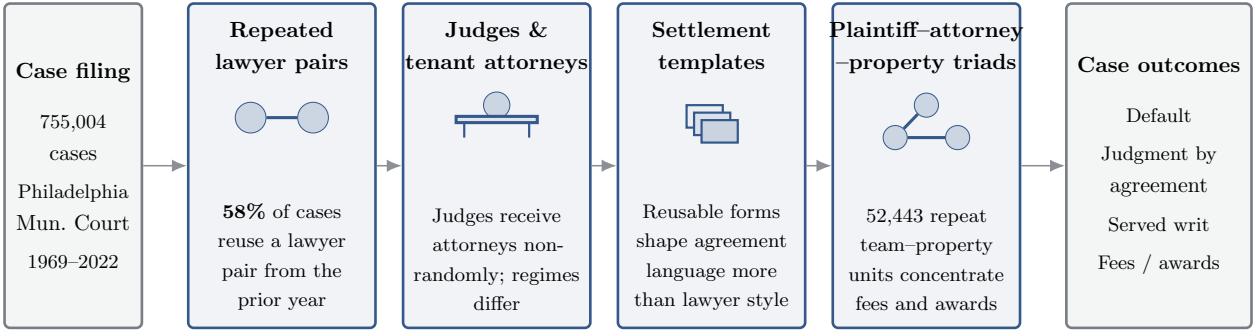

The four subsections that follow correspond to the four pillars of \figref{fig:framework}: repeated lawyer pairs, judge and tenant-attorney regimes, settlement templates, and plaintiff--attorney--property triads.

\subsection{Repeated attorney pairs structure the represented courtroom field}\label{sec:results_pairs}

The represented legal field is highly relational. In the both-represented and both-attorney-named sample ($n = 26{,}259$), 58.2\% of cases share the same plaintiff-attorney $\times$ tenant-attorney pairing as a case from the prior 12 months, and the mean prior same-pair count is 7.20. Repetition is the modal condition under which represented cases are processed, not a marginal tail.

That repetition tracks a distinct outcome ladder. With no prior same-pair exposure in the preceding year, the default rate is 17.9\%, the JBA rate is 52.9\%, and the served-writ rate is 17.8\%. At the highest repeat bucket (six or more prior encounters), the default rate falls to 9.7\%, the JBA rate rises to 72.1\%, and the served-writ rate rises to 26.3\%. Repeated lawyer--lawyer encounters thus shift cases away from default and toward negotiated resolution while preserving a substantial downstream enforcement channel.

\begin{figure}[H]
    \centering
    \includegraphics[width=0.72\linewidth]{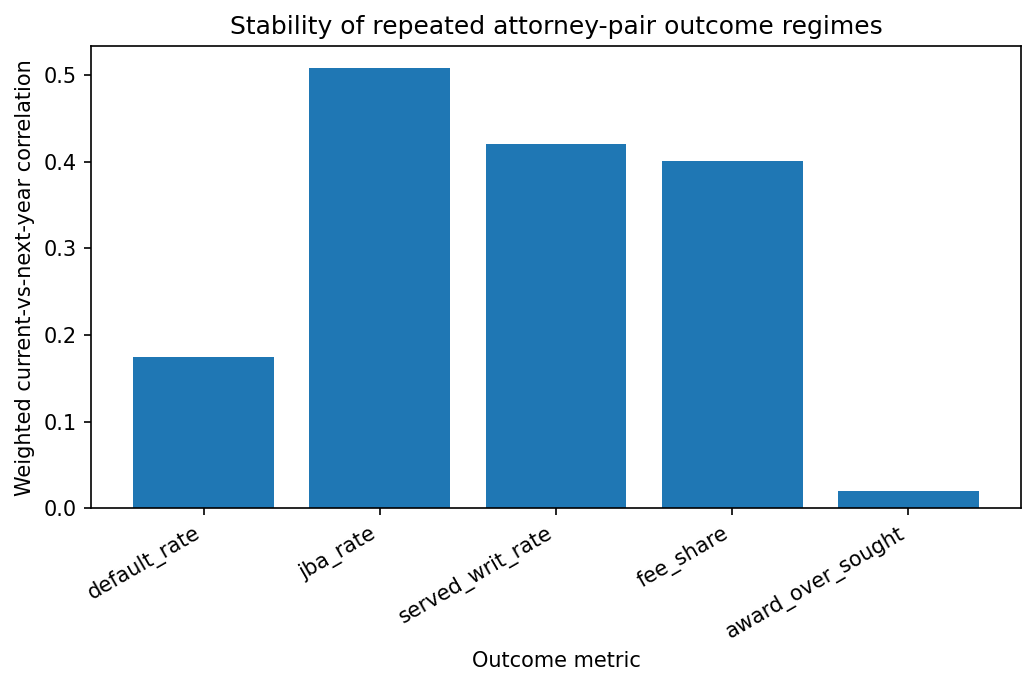}
    \caption{Stability of repeated attorney-pair outcome regimes. Repeated plaintiff-attorney $\times$ tenant-attorney pairings reproduce recognizable outcome patterns over time, especially in judgment by agreement, served writ, and fee share.}
    \label{fig:pair_stability}
\end{figure}

Regression estimates corroborate the descriptive pattern. Each additional prior same-pair encounter is associated with a 1.5 percentage-point decrease in default ($\beta=-0.0150$, $p<0.001$), a 5.1-point increase in judgment by agreement ($\beta=0.0506$, $p<0.001$), and a 1.5-point increase in served writ ($\beta=0.0149$, $p=0.001$). Repeated pair structure is also stable over time: across pair-years, the weighted current-to-next-year correlation equals 0.174 for default, 0.507 for JBA, 0.421 for served writ, and 0.401 for fee share, while award-over-sought is much less stable at 0.020 (\figref{fig:pair_stability}). The pair field reproduces recognizable regimes over time, especially in bargaining and enforcement rather than in award inflation.

The repeated-pair pattern requires careful interpretation. In pooled omnibus tests with year fixed effects alone, pair-specific structure is significant for default ($\eta^2 = 0.0822$, $p < 10^{-23}$), JBA ($\eta^2 = 0.1374$, $p < 10^{-110}$), served writ ($\eta^2 = 0.0957$, $p < 10^{-40}$), fee share ($\eta^2 = 0.1088$, $p < 10^{-28}$), and award-over-sought ($\eta^2 = 0.1197$, $p < 10^{-39}$). Once judge, plaintiff-attorney, and tenant-attorney main effects are absorbed simultaneously, the residual pair pattern attenuates sharply. What appears to be dyadic structure is therefore largely carried by repeated attorney identities and the courtroom environments into which those attorneys sort, rather than by a universal pair-specific interaction effect.

The represented legal field is also connected without being fully integrated. Attorney-name overlap across plaintiff-side and tenant-side roles equals 807 names, well below the 1{,}179.5 names implied by a fully integrated bar benchmark (hypergeometric $p<10^{-295}$). At the same time, same-name same-case overlap exceeds what would be expected under random within-year pairing (826 observed versus 291.3 expected, permutation $p=0.005$). The plaintiff-side and tenant-side bars therefore overlap, but they are not one unified market.

A final qualification concerns bargaining-text claims. Pair-identified JBA text is too sparse to support stable pair-level contract-style inference: most attorney pairs appear only once, and the in-sample fit of pair-mean strictness models collapses under holdout evaluation. We therefore exclude pair-level JBA-text fit from the substantive claims. Table~\ref{tab:pair_field} summarizes the core pair results; \subsecref{app:A2} reports the support and holdout diagnostics. The strongest pair evidence lies in routinized outcome regimes, not in pair-specific contract-style prediction.

\begin{table}[H]
\centering
\caption{Courtroom field and repeated attorney-pair structure: selected core estimates.}
\label{tab:pair_field}
\small
\begin{tabular}{p{0.39\linewidth}p{0.20\linewidth}p{0.15\linewidth}p{0.18\linewidth}}
\toprule
Finding & Estimate & Inference & Interpretation \\
\midrule
Cases with prior same attorney pair within 12 months & 0.582 & descriptive & repeated attorney encounters are the modal represented condition \\
Mean prior same-pair count & 7.20 & descriptive & repeated pairs are not only common, but often repeated many times \\
Default / JBA / served writ at pair bucket 0 & 0.179 / 0.529 / 0.178 & descriptive & non-recurring pairs are more default-heavy and less agreement-heavy \\
Default / JBA / served writ at pair bucket 6+ & 0.097 / 0.721 / 0.263 & descriptive & dense pair repetition shifts cases toward agreement and enforcement \\
Prior pair exposure $\rightarrow$ default / JBA / served writ & $-0.0150$ / $+0.0506$ / $+0.0149$ & all $p\leq 0.001$ & more prior pair exposure changes case flow \\
Pair omnibus test (year FE): default / JBA / served writ & $\eta^2 = 0.0822$ / $0.1374$ / $0.0957$ & all $p<10^{-23}$ & pair-specific regimes are descriptively substantive \\
Observed cross-side attorney-name overlap & 807 vs.\ 1{,}179.5 expected & $p<10^{-295}$ & the two bars overlap, but are not one integrated market \\
Same-name same-case overlap & 826 vs.\ 291.3 expected & $p=0.005$ & some cross-side overlap remains concentrated in actual case encounters \\
Pair-identified JBA text support & 121 texts / 106 pairs & diagnostic & pair-level bargaining-text claims face severe support constraints \\
Pair strictness model: in-sample vs.\ holdout & $R^2=0.892$ vs.\ $-0.908$ & diagnostic & the in-sample pair-text fit is overfit, not durable evidence \\
\bottomrule
\end{tabular}
\end{table}

\subsection{Judges and tenant-side attorneys structure the courtroom field}\label{sec:results_judges}

Observed case outcomes and actor mixes vary across judge-linked cases. We first characterize who appears before whom. In the judge-linked residential sample, top-\(k\) association tests on the 25 most frequent actors in each role (to reduce sparse-cell problems) show that judge \(\times\) plaintiff-attorney and judge \(\times\) tenant-attorney pairings deviate sharply from independence: \(\chi^2 = 7{,}576.2\) (df = 576) for judge \(\times\) plaintiff attorney and \(\chi^2 = 2{,}301.3\) (df = 576) for judge \(\times\) tenant attorney (both \(p<0.001\)). Actor mixes are not evenly distributed across judge-linked cases. The broader judge \(\times\) plaintiff statistic is sensitive to the plaintiff-key convention and is therefore not used as a replacement estimate.

Judge-specific baseline regimes match the sorting pattern. After residualizing month structure, pooled judge ANOVA tests remain significant for default (\(F=3.72\), \(\eta^2=0.0043\)), JBA (\(F=6.55\), \(\eta^2=0.0076\)), served writ (\(F=6.05\), \(\eta^2=0.0070\)), continuances (\(F=5.73\), \(\eta^2=0.0066\)), fee share (\(F=6.98\), \(\eta^2=0.0087\)), and award-over-sought (\(F=6.68\), \(\eta^2=0.0216\)); all six \(p\)-values are below the \(10^{-15}\) software floor. Judge-linked cases do not form identical baseline settings even after temporal adjustment.

The courtroom field is not organized by judges alone; tenant-attorney identity also explains meaningful variance. In broad month-fixed-effect tests, tenant-attorney identity is significant for default (\(\eta^2=0.0358\)), JBA (\(\eta^2=0.0578\)), served writ (\(\eta^2=0.0391\)), fee share (\(\eta^2=0.0458\)), award-over-sought (\(\eta^2=0.0347\)), and JBA strictness (\(\eta^2=0.0660\)), all with \(p<0.001\). Under the tightest plaintiff--judge--month specification, default remains significant (\(\eta^2=0.0635\), \(p=0.0046\)). Broad specifications thus reveal substantial bargaining and enforcement heterogeneity across tenant counsel; the tightest judge-linked comparison leaves default as the most robust tenant-attorney signal.

An assignment-cell audit sharpens this interpretation. Candidate cells with meaningful support fail balance: courtroom-month cells cover 63.8\% of judge-linked cases but yield a minimum balance \(p\)-value of \(4.1\times10^{-30}\); courtroom-week cells cover 61.3\% with minimum balance \(p=2.4\times10^{-16}\); courtroom-day cells cover 47.6\% with minimum balance \(p=2.9\times10^{-11}\). The representation-specific day cell achieves the best balance profile after multiple-testing adjustment, yet covers only 29.6\% of judge-linked cases with a median of three cases per usable cell. The docket does not sustain a broad clean judge lottery; judges and courtrooms function as institutional regimes, not as randomized judicial treatments.

A separate support diagnostic rules out a stable judge \(\times\) triad interaction. The 1{,}523 judge \(\times\) triad cells have a maximum size of four, with no cell reaching five cases; no outcome-specific judge \(\times\) triad model is estimable at a minimum threshold of five. This is a support failure, not a null treatment effect. The data describe judge regimes, attorney sorting, and triad structure, but cannot identify judge \(\times\) triad causal effects.

\begin{table}[H]
\centering
\caption{Courtroom governance through judges and tenant attorneys: selected statistically supported estimates and diagnostics. Reported \(p<10^{-15}\) values reflect the software floor for the underlying numerical routine; exact \(p\)-values are smaller.}
\label{tab:judge_tenant}
\small
\begin{tabular}{p{0.39\linewidth}p{0.20\linewidth}p{0.15\linewidth}p{0.18\linewidth}}
\toprule
Finding & Estimate & Inference & Interpretation \\
\midrule
Judge sorting with plaintiff attorneys / tenant attorneys & \(\chi^2 = 7{,}576.2\) / \(2{,}301.3\) & both \(p<0.001\) & courtroom actor mixes are relationally structured \\
Judge regime test: default / JBA / served writ & \(\eta^2=0.0043\) / \(0.0076\) / \(0.0070\) & all \(p<10^{-15}\) & judges differ in baseline case regimes \\
Judge regime test: continuances / fee share / award-over-sought & \(\eta^2=0.0066\) / \(0.0087\) / \(0.0216\) & all \(p<10^{-15}\) & judges also differ in procedural and monetary regimes \\
Tenant-attorney broad outcome regime: default / JBA / served writ & \(\eta^2=0.0358\) / \(0.0578\) / \(0.0391\) & all \(p<0.001\) & tenant-side legal intermediaries are not interchangeable \\
Tenant-attorney tightest FE: default & \(\eta^2=0.0635\) & \(p=0.0046\) & default remains the robust tight-specification tenant-attorney signal \\
Assignment-cell audit & broad cells fail balance & min \(p<10^{-10}\) & no broad clean judge lottery \\
Judge \(\times\) triad support & max group size \(=4\); groups \(n\ge5=0\) & diagnostic & judge--triad interactions are not estimable \\
\bottomrule
\end{tabular}
\end{table}

\subsection{Bargaining is standardized through dispersed templates}\label{sec:results_bargaining}

Settlement is not a procedural afterthought; it constitutes a contractual layer with its own architecture. Across 220{,}622 JBA texts, deadline language appears in 26.3\%, move-out language in 13.9\%, payment-plan language in 15.1\%, waiver clauses in 5.3\%, and lockout-trigger language in 9.6\%, yielding a mean strictness score of 0.742 on the 0--5 index (i.e., the average agreement contains well under one of the five clause types). Agreement content is a large, structured layer of case processing, not a small textual tail of the docket.

Raw attorney effects on that layer are nontrivial, but templates dominate. In the expanded template decomposition, raw attorney $\eta^2$ equals 0.156 for strictness, 0.492 for waiver, 0.337 for lockout-trigger language, 0.144 for payment-plan language, 0.116 for move-out language, 0.186 for deadline language, and 0.735 for time-is-essence language. Raw template $\eta^2$ is substantially larger across nearly every family: 0.909 for strictness, 0.879 for waiver, 0.900 for lockout triggers, 0.971 for payment plans, 0.977 for move-out language, 0.961 for deadline language, and 0.882 for time-is-essence language. Once template fixed effects are absorbed, attorney $\eta^2$ collapses to roughly 0.003--0.005 across clause families. The contractual layer is better described as template governance---form-document reuse and bargaining through standardized scripts---than as generic attorney style.

\begin{figure}[H]
    \centering
    \includegraphics[width=0.82\linewidth]{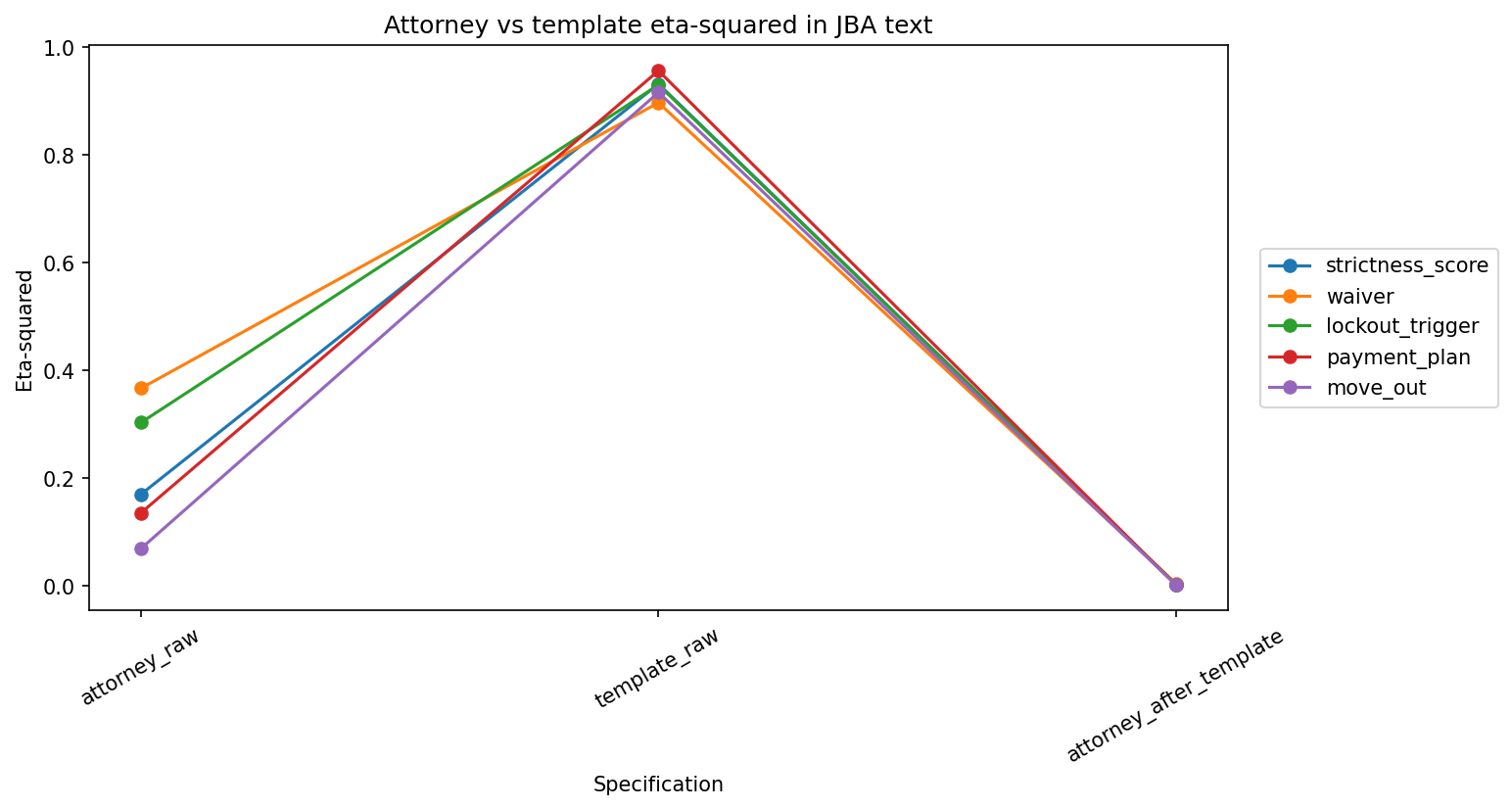}
    \caption{Attorney versus template governance in JBA text. Raw attorney effects exist, but template-level structure is much larger, and attorney explanatory power collapses after template fixed effects are absorbed.}
    \label{fig:template_eta2}
\end{figure}

Template dominance does not imply that a single boilerplate governs the whole court. The top five templates account for 7.4\% of deadline clauses, 3.4\% of move-out clauses, 2.1\% of waiver clauses, 1.4\% of lockout-trigger clauses, and 1.2\% of payment-plan clauses. Bargaining is standardized through many reusable forms, not through one universal document. At the same time, individual attorneys rely heavily on a small part of that menu. For the most-used deadline template, 99.7\% of texts come from a single attorney; for the most-used lockout-trigger template, 98.6\% come from a single attorney (the ``top template's dominant-attorney share''; Table~\ref{tab:template_appendix}). This is concentration of a specific template within one attorney, not of an attorney's practice within one template.

Tenant-attorney identity also shapes the contractual layer in both the broad and within-plaintiff-month specifications. Under within-plaintiff-month controls, tenant-attorney identity remains significant for strictness ($\eta^2=0.0498$, $p<10^{-5}$), move-out language ($\eta^2=0.0430$, $p<0.001$), and lockout-trigger language ($\eta^2=0.0423$, $p<0.001$), with weaker evidence for waiver and payment-plan language. The contractual layer is therefore not governed by plaintiff-side forms alone; the tenant-side legal field that encounters those forms also shapes their content. The strictest plaintiff--judge--month text specifications are too thin to estimate.

Stricter text does not map mechanically onto more visible enforcement. Each one-unit increase in the strictness index (range 0--5) is associated with a 1.8 percentage-point decrease in same-address refiling within 12 months ($\beta=-0.0180$, $p<10^{-34}$), a 1.5-point decrease in served writ ($\beta=-0.0151$, $p<10^{-30}$), and an 8.8-point decrease in recorded breach ($\beta=-0.0882$, $p<0.001$), but a 0.6-point increase in cases lacking a recorded satisfaction entry ($\beta=0.0056$, $p<10^{-6}$). Stricter agreements structure future obligations and enforcement possibilities non-monotonically: more restrictive text does not translate uniformly into more visible downstream enforcement.

\begin{table}[H]
\centering
\caption{Bargaining, templates, and downstream agreement governance: selected core estimates.}
\label{tab:bargaining}
\small
\begin{tabular}{p{0.39\linewidth}p{0.20\linewidth}p{0.15\linewidth}p{0.18\linewidth}}
\toprule
Finding & Estimate & Inference & Interpretation \\
\midrule
JBA text prevalence: deadline / move-out / payment plan / waiver / lockout & 0.263 / 0.139 / 0.151 / 0.053 / 0.096 & descriptive & settlement contains a large contractual layer \\
Template raw $\eta^2$: strictness / waiver / lockout & 0.909 / 0.879 / 0.900 & all $p<0.001$ & reusable textual forms dominate raw attorney effects \\
Template raw $\eta^2$: payment / move-out / deadline / time-is-essence & 0.971 / 0.977 / 0.961 / 0.882 & all $p<0.001$ & template dominance extends across additional clause families \\
Attorney $\eta^2$ after template FE & about 0.003--0.005 & small & most raw attorney style collapses after template absorption \\
Top-5 template share of clause-positive texts & 0.012--0.074 & descriptive & bargaining is standardized through a menu of forms, not one monopoly text \\
Dominant attorney's share of dominant deadline / lockout template & 0.997 / 0.986 & descriptive & some template governance is field-dispersed but attorney-carried in practice \\
Tenant-attorney within-plaintiff-month effect on strictness / move-out / lockout & $\eta^2 = 0.0498$ / $0.0430$ / $0.0423$ & all $p<0.001$ & tenant-side counsel also shapes contract terms \\
Strictness $\rightarrow$ same-address refiling / served writ & $-0.0180$ / $-0.0151$ & both $p<10^{-30}$ & stricter agreements do not simply raise visible downstream enforcement \\
Strictness $\rightarrow$ no recorded satisfaction & $+0.0056$ & $p<10^{-6}$ & harsher agreements are associated with more unresolved monetary closure \\
Strictness $\rightarrow$ recorded breach & $-0.0882$ & $p<0.001$ & contract severity and observed breach are not a simple monotone relationship \\
\bottomrule
\end{tabular}
\end{table}

\subsection{Fees and awards are organized through repeated teams and places}\label{sec:results_debt}

The monetary layer of eviction court concentrates in repeated operational units. In the fee-share decomposition, the plaintiff--attorney--property triad carries the largest effect size (\(\eta^2=0.0846\), \(p<10^{-240}\)), followed by address-level and plaintiff--attorney structure. The same pattern holds for award-over-sought, where the triad again leads (\(\eta^2=0.0999\), \(p<0.001\)). The triad functions as a composite unit in which legal representation, plaintiff identity, and property recur jointly across cases.

The canonical repeat-triad typology contains 52{,}443 repeated plaintiff--attorney--property units in the named-case universe. Most repeat-triad cases sit inside broader plaintiff--attorney portfolios rather than within a single address: the portfolio-embedded weighted share is 0.582, the address-captured weighted share is 0.165, the median plaintiff--attorney pair has seven addresses, and the median repeat triad accounts for 40\% of cases at its address. The remaining 25.3\% of repeat-triad cases are neither portfolio-embedded nor address-captured under the present definitions; this residual reflects intermediate cases that partially satisfy both criteria without crossing the classification thresholds. We treat this as the substantive triad universe; a separate integration check yields a narrower diagnostic universe of 2{,}712 groups, which we reserve for diagnostic support checks rather than as a replacement typology.

Case-level models sharpen the debt pattern. Each additional prior same-plaintiff--attorney encounter is associated with a 0.19 percentage-point increase in fee share ($\beta=0.00185$, $p<10^{-10}$) and a 2.5-point decrease in award-over-sought ($\beta=-0.0251$, $p<10^{-45}$). Generic repeat-address exposure also lowers award-over-sought ($\beta=-0.0211$, $p<10^{-6}$), but same-plaintiff same-address repetition raises it ($\beta=0.0280$, $p<10^{-8}$). Repeated legal teams add fee burden inside the case-processing channel, while repeated place-specific filing tracks award inflation.

Cases with higher monetary burden display a distinctive outcome profile. A 1.0-unit increase in fee share (i.e., moving from no fees to fees equal to the entire award) is associated with a 34.3 percentage-point increase in default ($\beta=0.343$, $p<0.001$) and a 58.1-point decrease in JBA ($\beta=-0.581$, $p<0.001$); a 1.0-unit increase in award-over-sought is associated with a 7.6-point increase in default ($\beta=0.0756$, $p<0.001$), a 6.8-point decrease in JBA ($\beta=-0.0681$, $p<0.001$), and a 9.7-point increase in served writ ($\beta=0.0970$, $p<0.001$). Monetary burden is part of courtroom governance but is not equivalent to physical possession enforcement.

\begin{table}[H]
\centering
\caption{Debt production through repeated operational units: selected statistically supported estimates.}
\label{tab:debt}
\small
\begin{tabular}{p{0.39\linewidth}p{0.20\linewidth}p{0.15\linewidth}p{0.18\linewidth}}
\toprule
Finding & Estimate & Inference & Interpretation \\
\midrule
Fee-share driver with largest effect size & triad \(\eta^2 = 0.0846\) & \(p<10^{-240}\) & repeated team + property is the strongest debt unit \\
Award-over-sought driver with largest effect size & triad \(\eta^2 = 0.0999\) & \(p<0.001\) & repeated team + property also dominates award inflation \\
Address effect on fee share / award-over-sought & \(\eta^2 = 0.0387\) / \(0.0312\) & both significant & place matters independently of individual actors \\
Canonical repeat triads & 52{,}443 & descriptive & common repeated team--property units \\
Repeat-triad weighted share: portfolio-embedded / address-captured & 0.582 / 0.165 & descriptive & most repeat triads implement broader servicing relations \\
Raw triad \(\eta^2\) on default / JBA / served writ & 0.0725 / 0.0990 / 0.0773 & all \(p<0.001\) & triads are strong raw operational units \\
Repeat plaintiff--attorney intensity \(\rightarrow\) fee share / award-over-sought & \(+0.00185\) / \(-0.0251\) & both highly significant & repeated teams add fee burden but are associated with lower award-over-sought \\
Same-plaintiff same-address repetition \(\rightarrow\) award-over-sought & \(+0.0280\) & \(p<10^{-8}\) & repeated place capture is closely tied to final award inflation \\
Fee share \(\rightarrow\) default / JBA & \(+0.343\) / \(-0.581\) & both \(p<0.001\) & debt-heavy cases are more adverse and less settlement-heavy \\
\bottomrule
\end{tabular}
\end{table}

\section{Discussion, conclusion, and limitations}\label{sec:discussion}

\subsection{Principal contribution}

Post-filing eviction case processing is organized through repeated institutional routines. Four linked features structure that governance in the Philadelphia docket. First, represented eviction is routinized through repeated plaintiff-attorney by tenant-attorney encounters rather than dominated by non-recurring cases. Second, judges and tenant attorneys structure baseline outcome, continuance, fee, award, and agreement regimes, but assignment diagnostics caution against reading these patterns as evidence of a broad randomized judge-lottery design. Third, settlement is a contractual layer governed primarily by reusable templates rather than by free-form individualized drafting. Fourth, fee burden and award inflation are organized most strongly through repeated team--property units.

Diagnostics clarify the boundary between statistical structure and causal identification. Candidate judge-assignment cells face a support--balance tradeoff; judge--triad interactions are not estimable because no judge--triad cell reaches five cases; and the Callaway--Sant'Anna recurrence analysis is informative only as a supplemental event diagnostic. Our claims describe a stable institutional field, with bounded diagnostics reported where the data permit them.

Together, these findings reframe the court record. Eviction court does not merely register housing disputes after filing; it organizes how those disputes are bargained, which texts are reused, how monetary burden is layered onto possession claims, and which repeated legal relationships structure case flow. The institutional process that follows filing is itself the object of analysis.

\subsection{Policy implications}

Policy intervention need not focus exclusively on landlords, tenants, or final judgments. If repeated attorney pairings structure case flow, then courtroom workgroups matter. If judges and tenant attorneys differ in baseline regimes, courtroom administration and tenant-side legal capacity matter. If settlement terms are template-governed, agreement text should be a policy object rather than an opaque procedural artifact. If debt production is strongest in repeated team--property units, fee regulation and monetary monitoring should be treated as housing-stability policy. The assignment and support diagnostics add a further implication: judge-linked patterns should be monitored as institutional regimes, but the docket alone should not be treated as a clean randomized judge experiment.

\subsection{Epistemic scope and limitations}

We do not claim strong causal identification for judges, attorney pairs, templates, clauses, or triads: none of these can be treated as a randomly assigned intervention in this docket. Judge sorting is documented as a structural feature of the courtroom field, not characterized as bias through a random-assignment design. Repeated-pair results describe routinized relationships and the regimes they reproduce, but residual dyadic effects weaken after attorney identities and courtroom sorting are absorbed. Contractual results show how settlement text is organized, but they do not turn individual templates or clauses into clean causal interventions. Triad results show where debt and repeated operational structure are concentrated, but triads are not randomly assigned.

Six data limitations follow. First, key identifiers are not equally informative across the full archival period. Second, pair-specific JBA text is sparse, motivating our reliance on support and holdout diagnostics before making pair-text claims. Third, fee fields are substantively informative only in the modern period. Fourth, JBA text is observed for a large but selected subset of negotiated cases. Fifth, administrative court records miss informal displacement and may contain party, address, or text-recording errors. Sixth, the separate integration check uses a narrower repeat-triad construction than the canonical typology; we therefore retain the 52{,}443-triad typology as the substantive Section~4.4 universe and reserve the narrower construction for diagnostics.

\subsection{Conclusion}

Eviction court is not only a venue where individual disputes are closed. In Philadelphia, it is a repeated institutional process organized by recurring lawyers, judge and tenant-attorney regimes, standardized settlement forms, and team--property routines that add fees and awards. Studying eviction after filing reframes the object of analysis: the central question is not only who files or who is removed, but how courtroom routines shape negotiation, written obligations, debt, and enforcement. Eviction court governs through repeat relationships, forms, and operational units; assignment and recurrence diagnostics mark the limits of causal interpretation.

\section*{Data and Code Availability}
The Philadelphia Municipal Court dataset used in this study is available from Philadelphia Legal Assistance at \url{https://docs.philalegal.org/index.php/s/w9IQZrb8eDqXJkU}. Replication code for data cleaning, measure construction, tables, and figures is available at \url{https://github.com/MariosPapamix}.

\bibliographystyle{apalike}
\bibliography{Eviction}

\clearpage
\appendix
\section*{Supplementary Information}
The supplement reports diagnostics for the courtroom-governance analyses: attorney-pair text support and holdout tests, judge-assignment support audits, judge and tenant-attorney heterogeneity, bargaining-template diagnostics, and fee/debt diagnostics.

\section{Pair-support diagnostics}\label{app:A}

\subsection{Pair-text support and holdout diagnostics}\label{app:A2}

\begin{figure}[H]
    \centering
    \includegraphics[width=0.80\linewidth]{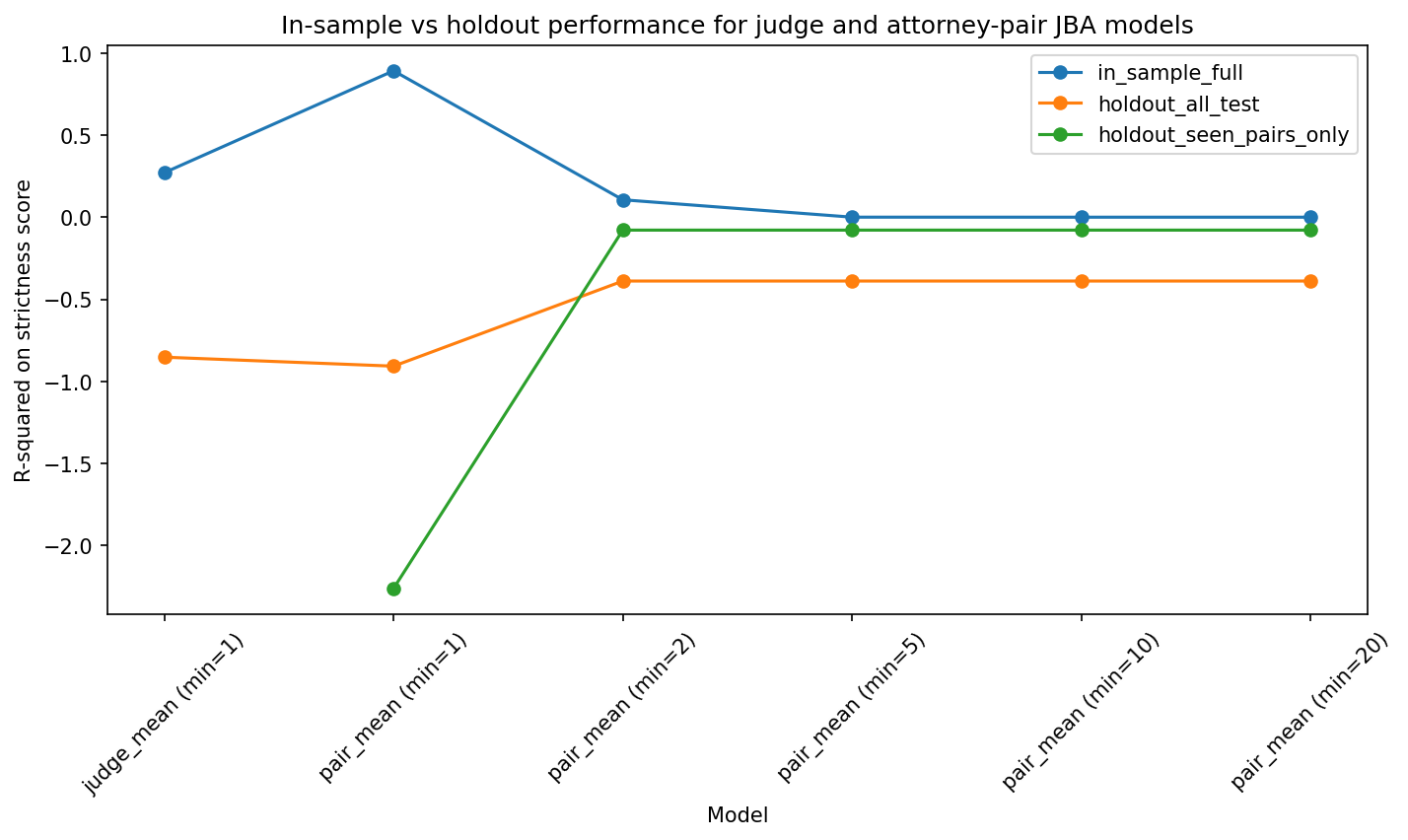}
    \caption{In-sample versus holdout performance for judge and attorney-pair JBA models. The apparent pair advantage is an in-sample artifact of support sparsity rather than durable predictive structure.}
    \label{fig:pair_holdout}
\end{figure}

The support problem in pair-identified JBA text warrants direct documentation. Table~\ref{tab:pair_support} reports the core support counts, and Figure~\ref{fig:pair_holdout} contrasts in-sample and holdout performance for judge and attorney-pair JBA models.

\begin{table}[H]
\centering
\caption{Pair-identified JBA-text support diagnostics.}
\label{tab:pair_support}
\small
\begin{tabular}{p{0.52\linewidth}p{0.20\linewidth}p{0.18\linewidth}}
\toprule
Diagnostic & Estimate & Interpretation \\
\midrule
Text rows with identified attorney pairs & 121 & extremely limited support universe \\
Unique attorney pairs & 106 & almost one pair per text row \\
Median texts per pair & 1.0 & pair support is typically singleton \\
75th / 90th percentile texts per pair & 1.0 / 2.0 & even the upper tail has little repeated support \\
Singleton-pair share & 0.887 & most pairs appear once \\
Share of pairs with fewer than 5 texts & 0.991 & almost no pair has meaningful repeated text support \\
In-sample judge-mean $R^2$ & 0.271 & modest baseline fit \\
In-sample pair-mean $R^2$ (min support = 1) & 0.892 & large fit driven by support sparsity \\
In-sample pair-mean $R^2$ (min support = 2) & 0.106 & fit collapses once singleton pairs are excluded \\
Holdout all-test pair-mean $R^2$ (min support = 1) & -0.908 & negative out-of-sample performance \\
Holdout seen-pairs-only pair-mean $R^2$ (min support = 1) & -2.262 & pair model fails even where the pair is seen in training \\
\bottomrule
\end{tabular}
\end{table}

\section{Supplemental courtroom-governance tests}\label{app:B}

\subsection{Calendar-cell audit for judge-assignment designs}

A supplementary audit evaluates whether narrower hearing-calendar cells approximate local quasi-random judicial variation. The exercise is informative but not decisive. Tighter cells can improve balance, but they do so by sharply reducing usable support. In the selected assignment-day design, 524 cells are usable, covering 28.6\% of eligible cases, with an average of 3.36 judges per usable cell. Placebo balance remains only moderate (mean absolute $t$-statistic $= 0.93$), and only 3 of 14 IV specifications clear a conventional first-stage threshold ($F \ge 10$). The audit therefore supports caution rather than a clean random-assignment interpretation.

\begin{figure}[H]
    \centering
    \begin{subfigure}[b]{0.48\linewidth}
        \includegraphics[width=\linewidth]{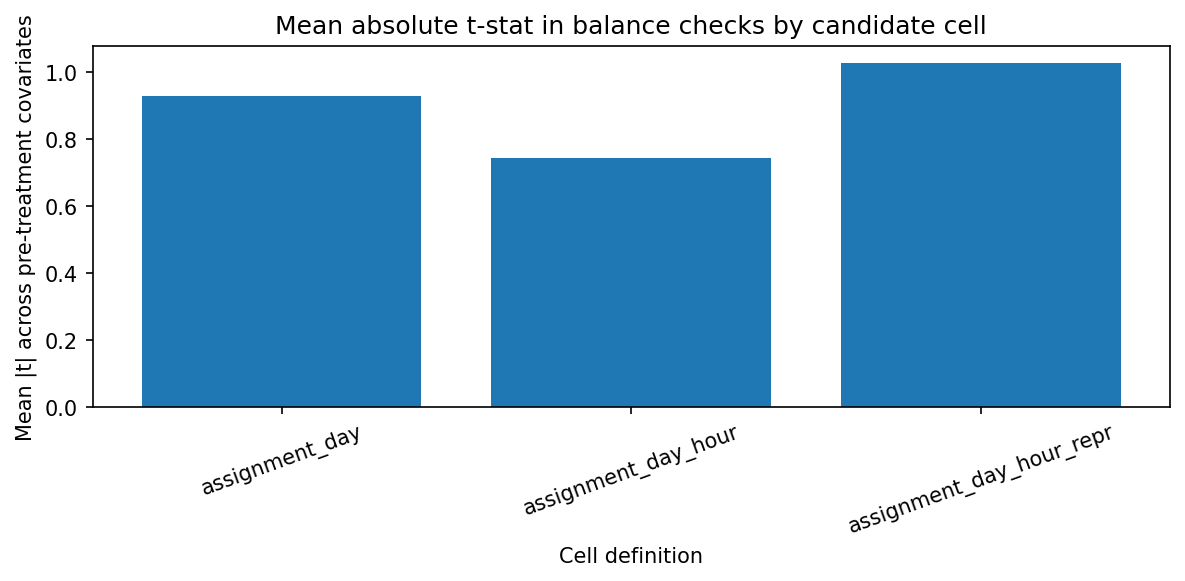}
        \caption{Mean absolute balance-test statistic by candidate cell.}
    \end{subfigure}
    \hfill
    \begin{subfigure}[b]{0.48\linewidth}
        \includegraphics[width=\linewidth]{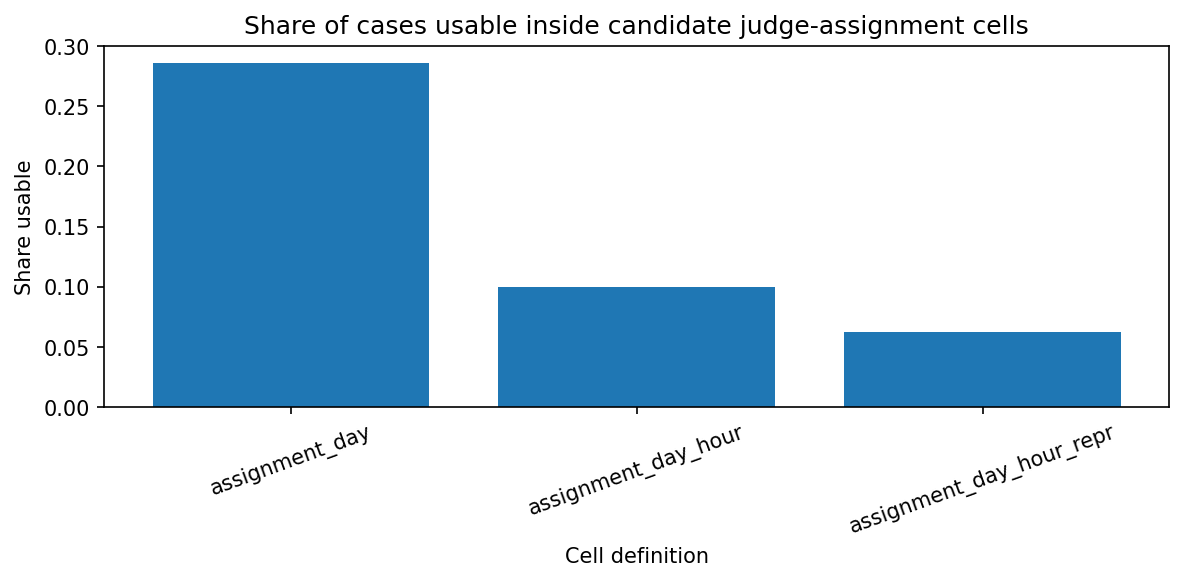}
        \caption{Share of usable cases by candidate cell.}
    \end{subfigure}
    \caption{Support--balance tradeoff in candidate judge-assignment cells. Narrower cells can improve comparability, but they sharply reduce usable support.}
    \label{fig:judge_cell_audit}
\end{figure}

\subsection{Judge sorting, judge regimes, and residual interaction}\label{app:B1}

Table~\ref{tab:judge_supp} reports the fuller judge-sorting and regime diagnostics that underwrite the main text. The central pattern is consistent across specifications: judges receive actor mixes non-randomly and differ in baseline case, continuance, fee, and award regimes, but the residual judge--attorney interaction is weak once judge and attorney main effects are absorbed.

\begin{table}[H]
\centering
\caption{Supplemental judge-sorting and judge-regime diagnostics.}
\label{tab:judge_supp}
\small
\begin{tabular}{p{0.44\linewidth}p{0.22\linewidth}p{0.12\linewidth}p{0.14\linewidth}}
\toprule
Finding & Estimate & $p$-value & Interpretation \\
\midrule
Judge $\times$ plaintiff-attorney sorting & $\chi^2 = 7{,}576.2$ (df = 576) & $<0.001$ & judges do not receive plaintiff attorneys randomly \\
Judge $\times$ plaintiff sorting & $\chi^2 = 14{,}154.2$ (df = 576) & $<0.001$ & plaintiffs are also sorted unevenly across judges \\
Judge $\times$ tenant-attorney sorting & $\chi^2 = 2{,}301.3$ (df = 576) & $<10^{-200}$ & tenant-side actors are unevenly distributed too \\
Judge regime test: default / JBA / served writ & $\eta^2 = 0.0063$ / $0.0140$ / $0.0108$ & all significant & judges differ in baseline case regimes \\
Judge regime test: continuances / fee share / award-over-sought & $\eta^2 = 0.0082$ / $0.0203$ / $0.0135$ & all significant & judges also differ in procedural and monetary regimes \\
Residual judge--attorney interaction: default / JBA / served writ & $F = 0.91$ / $1.11$ / $0.89$ & 0.847 / 0.125 / 0.894 & little residual interaction effect remains after fixed structures \\
Judge-only adjusted JBA strictness tests & weak after adjustment & diagnostic & not used as a substantive judge-centered contract claim \\
\bottomrule
\end{tabular}
\end{table}

\subsection{Tenant-attorney heterogeneity under broad and strict controls}\label{app:B2}

Table~\ref{tab:tenant_supp} shows that tenant-attorney heterogeneity survives stricter controls. The broad specifications produce larger estimates, but the stricter within-plaintiff-year and plaintiff--judge--event-month specifications continue to display meaningful tenant-attorney variation in visible case outcomes and in settlement content.

\begin{table}[H]
\centering
\caption{Tenant-attorney heterogeneity under broad and stricter controls.}
\label{tab:tenant_supp}
\small
\begin{tabular}{p{0.30\linewidth}p{0.16\linewidth}p{0.16\linewidth}p{0.14\linewidth}p{0.16\linewidth}}
\toprule
Outcome or contractual feature & Broad $\eta^2$ & Strict $\eta^2$ & Strict $p$-value & Interpretation \\
\midrule
Default & 0.0257 & 0.0136 & $<3\times10^{-12}$ & tenant-attorney variation survives within-plaintiff-year comparisons \\
Judgment by agreement & 0.0504 & 0.0209 & $<6\times10^{-32}$ & tenant-side heterogeneity is strongest in bargaining \\
Served writ & 0.0364 & 0.0160 & $<5\times10^{-18}$ & enforcement paths also vary across tenant counsel \\
Fee share & 0.0133 & 0.0100 & 0.007 & debt burden differs across tenant attorneys \\
Strictness score & 0.0640 & 0.0241 & $<10^{-20}$ & tenant attorneys also shape contract severity \\
Move-out clause & 0.0519 & 0.0219 & $<10^{-16}$ & exit language varies meaningfully \\
Waiver clause & 0.0176 & 0.0108 & 0.044 & waiver language also varies under strict controls \\
Lockout-trigger clause & 0.0341 & 0.0138 & $<10^{-4}$ & enforcement-trigger language differs across counsel \\
Payment-plan clause & 0.0269 & 0.0116 & 0.010 & installment structure varies too \\
\bottomrule
\end{tabular}
\end{table}

\section{Additional bargaining, template, and debt diagnostics}\label{app:C}

\subsection{Agreement-text prevalence and fee series}\label{app:C1}

\begin{figure}[H]
    \centering
    \begin{subfigure}[b]{0.48\linewidth}
        \includegraphics[width=\linewidth]{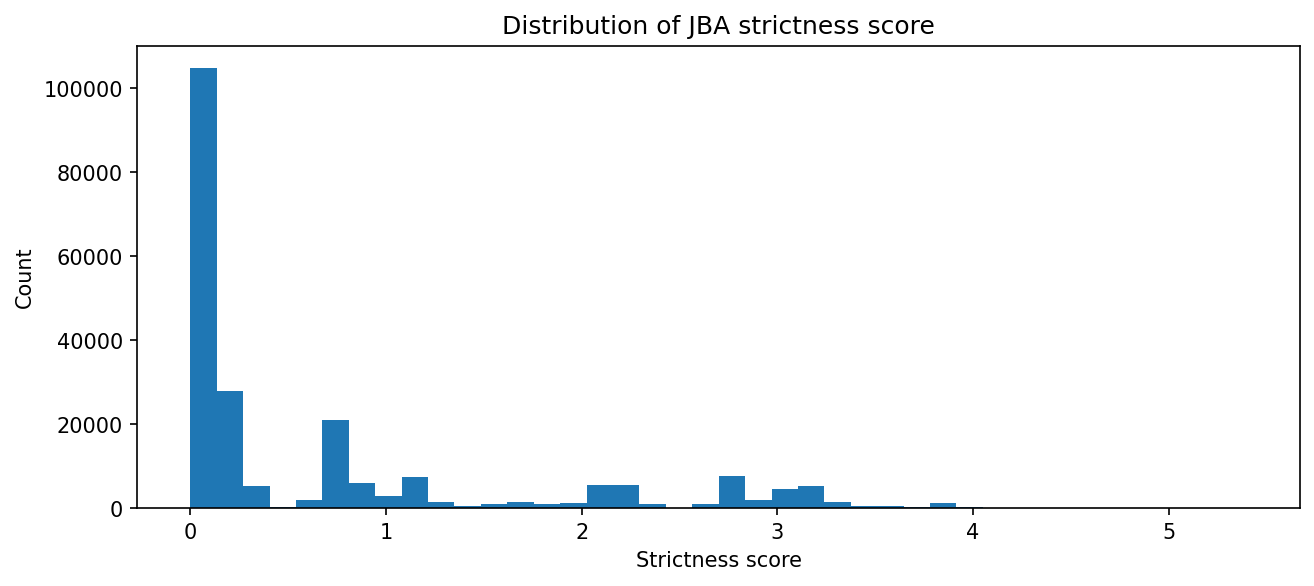}
        \caption{Distribution of JBA strictness.}
    \end{subfigure}
    \hfill
    \begin{subfigure}[b]{0.48\linewidth}
        \includegraphics[width=\linewidth]{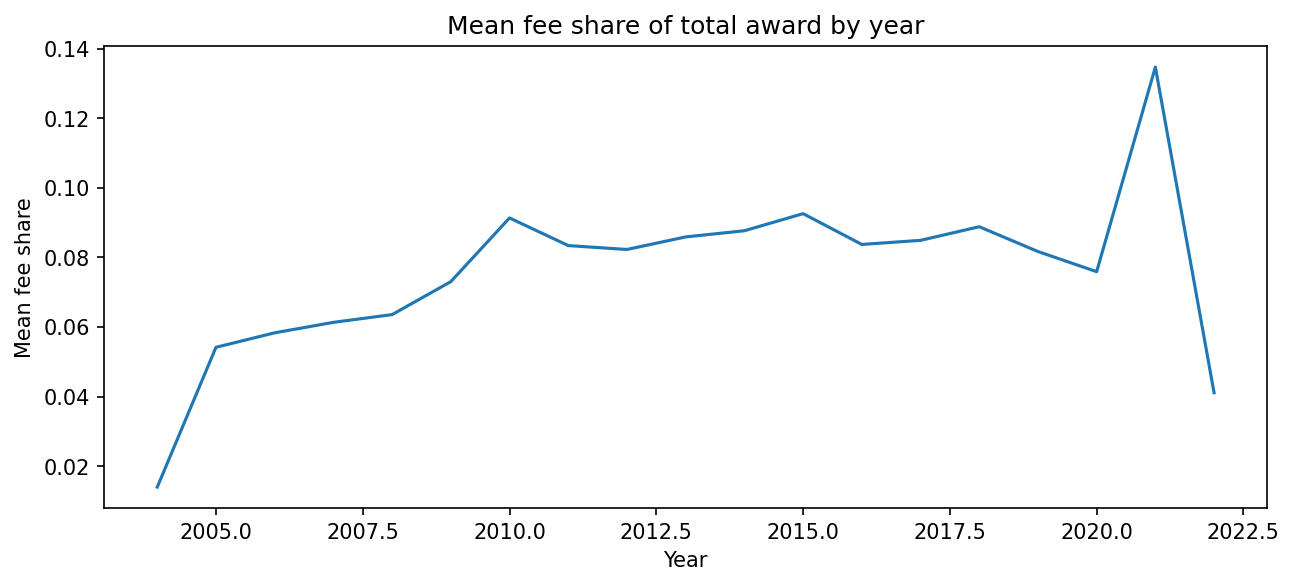}
        \caption{Mean fee share by year.}
    \end{subfigure}
    \caption{Additional bargaining and debt diagnostics. Agreement text has a broad strictness distribution, and fee share remains elevated but non-monotone in the modern period.}
    \label{fig:jba_fee_appendix}
\end{figure}

Figure~\ref{fig:jba_fee_appendix} visualizes the broad JBA strictness distribution and the non-monotone but elevated fee-share series in the modern period. Table~\ref{tab:jba_features_appendix} reports the core agreement-text prevalence and validity checks used in the analysis.

\begin{table}[H]
\centering
\caption{Selected JBA text features and validity checks.}
\label{tab:jba_features_appendix}
\small
\begin{tabular}{p{0.50\linewidth}p{0.18\linewidth}p{0.22\linewidth}}
\toprule
Feature or validity check & Estimate & Interpretation \\
\midrule
Non-empty JBA texts & 220{,}622 & descriptive text universe \\
Deadline language & 0.263 & explicit deadlines are common \\
Move-out language & 0.139 & explicit exit conditions are common \\
Payment-plan language & 0.151 & installment payment structure is common \\
Waiver language & 0.053 & waiver clauses are present but less common \\
Lockout-trigger language & 0.096 & enforcement triggers appear in a non-trivial share of agreements \\
Mean strictness score & 0.742 & agreements cluster at low but non-zero strictness \\
Additive versus PCA strictness correlation & 0.722 & strong alignment across constructions \\
PC1 explained variance & 0.306 & first component captures a meaningful share of clause variance \\
\bottomrule
\end{tabular}
\end{table}

\subsection{Template concentration and attorney template dependence}\label{app:C2}

\begin{figure}[H]
    \centering
    \includegraphics[width=0.72\linewidth]{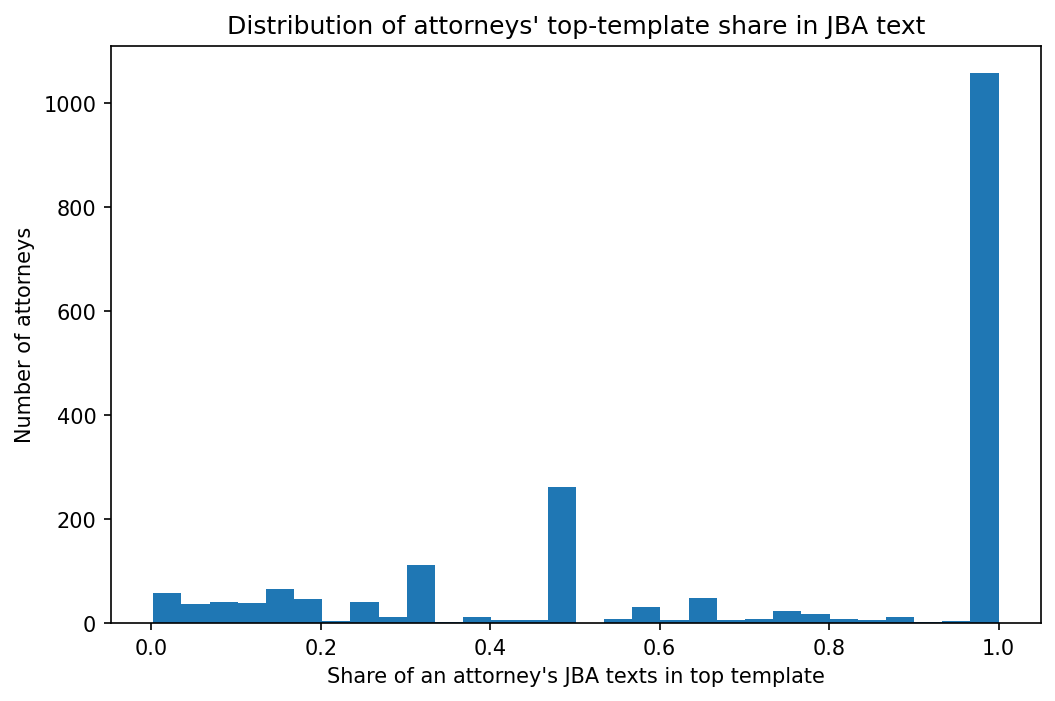}
    \caption{Attorney dependence on a top template. Attorneys vary in how heavily they rely on one form, but the distribution is broad rather than collapsed onto a single monopoly document.}
    \label{fig:template_dependence_appendix}
\end{figure}

Table~\ref{tab:template_appendix} reports the clause-family concentration details. Figure~\ref{fig:template_dependence_appendix} shows the distribution of each attorney's top-template share---the fraction of an attorney's own JBA texts that come from his or her single most-used template. This is the converse measure to the ``top template's dominant-attorney share'' reported in Table~\ref{tab:template_appendix}: the table asks how concentrated a given template is across attorneys, while the figure asks how concentrated each attorney is across templates.

\begin{table}[H]
\centering
\caption{Detailed clause-family template concentration.}
\label{tab:template_appendix}
\small
\begin{tabular}{p{0.24\linewidth}p{0.16\linewidth}p{0.16\linewidth}p{0.16\linewidth}p{0.18\linewidth}}
\toprule
Clause family & Positive texts & Largest template share & Largest-five share & Largest template's dominant-attorney share \\
\midrule
Deadline & 57{,}986 & 0.033 & 0.074 & 0.997 \\
Move-out & 30{,}601 & 0.012 & 0.034 & 0.500 \\
Payment plan & 33{,}378 & 0.003 & 0.012 & 0.505 \\
Waiver & 11{,}750 & 0.005 & 0.021 & 0.603 \\
Lockout trigger & 21{,}284 & 0.003 & 0.014 & 0.986 \\
\bottomrule
\end{tabular}
\end{table}

\subsection{Repeat-triad typology and fee-driver structure}\label{app:C3}

\begin{figure}[H]
    \centering
    \begin{subfigure}[b]{0.48\linewidth}
        \includegraphics[width=\linewidth]{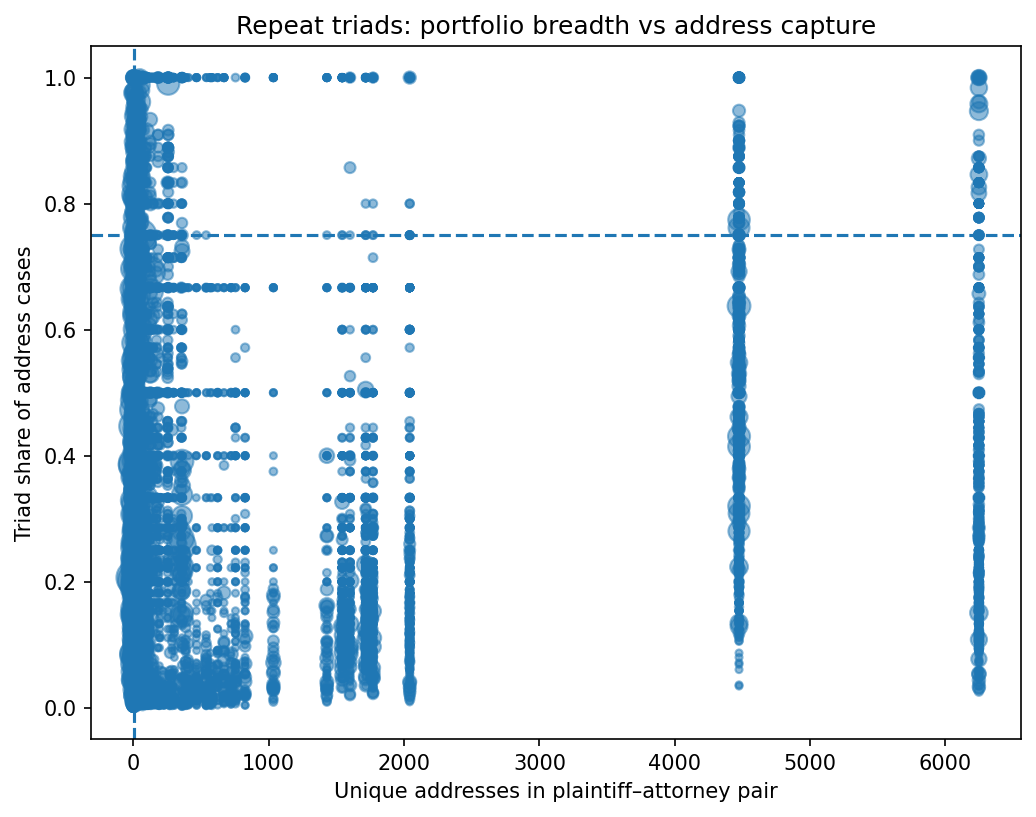}
        \caption{Portfolio embedding versus address capture.}
    \end{subfigure}
    \hfill
    \begin{subfigure}[b]{0.48\linewidth}
        \includegraphics[width=\linewidth]{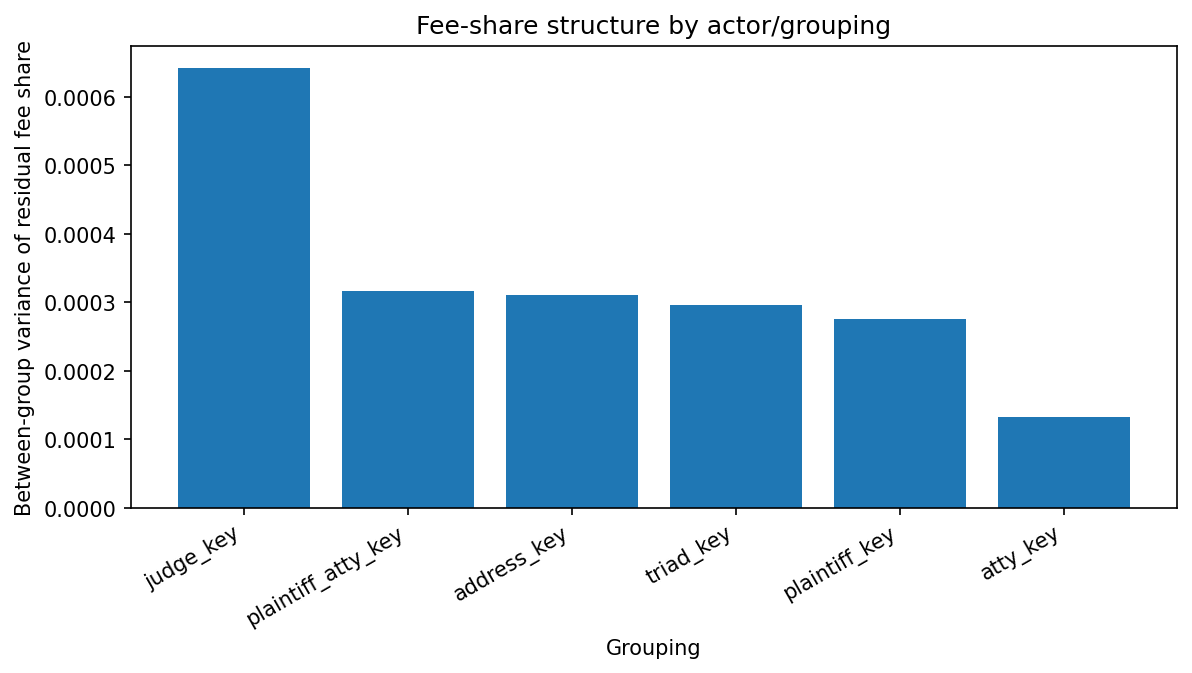}
        \caption{Fee-driver strength by actor and repeated unit.}
    \end{subfigure}
    \caption{Additional triad and fee-governance diagnostics. Repeat triads are more often portfolio-embedded than address-captured, and fee burden is most strongly structured at the triad and address levels.}
    \label{fig:triad_fee_appendix}
\end{figure}

The repeat-triad outputs show that the triad is a substantive operational unit, typically embedded in a broader servicing relationship rather than confined to a one-address pipeline. Table~\ref{tab:triad_appendix} reports the core typology statistics. Figure~\ref{fig:triad_fee_appendix} adds the visual diagnostics for triad typology and fee-driver strength.

\begin{table}[H]
\centering
\caption{Repeat-triad typology summary.}
\label{tab:triad_appendix}
\small
\begin{tabular}{p{0.54\linewidth}p{0.18\linewidth}p{0.18\linewidth}}
\toprule
Statistic & Estimate & Interpretation \\
\midrule
Repeat triads in the named-case universe & 52{,}443 & repeated team--property units are common \\
Portfolio-embedded weighted case share & 0.582 & most repeat-triad cases sit inside broader plaintiff--attorney portfolios \\
Address-captured weighted case share & 0.165 & a smaller share looks tied to one repeatedly used address \\
Median addresses per plaintiff--attorney pair among repeat triads & 7.0 & the median repeat triad sits inside a multi-address relation \\
Median triad share of cases at its address & 0.400 & the median triad accounts for 40\% of cases at its address \\
Weighted scripted share among repeat triads with 5+ JBA texts & 0.103 & only a minority of supported repeat triads exhibit clearly scripted patterns \\
\bottomrule
\end{tabular}
\end{table}

\section{Additional causal-boundary diagnostics}\label{app:D}

\subsection{Judge-assignment support and balance}\label{app:D1}

Table~\ref{tab:judge_lottery_appendix} reports the candidate judge-assignment-cell audit. Broad cells provide support but fail balance; the most restrictive representation-specific day cell improves the balance profile only by covering less than one-third of judge-linked cases. The main text therefore interprets judge-linked patterns as courtroom-regime evidence rather than as a clean judge lottery.

\begin{table}[H]
\centering
\caption{Judge/courtroom assignment-cell diagnostic: support and balance.}
\label{tab:judge_lottery_appendix}
\small
\begin{tabular}{p{0.28\linewidth}p{0.18\linewidth}p{0.18\linewidth}p{0.18\linewidth}p{0.12\linewidth}}
\toprule
Candidate cell & Usable share & Median cases per cell & Minimum balance $p$ & Verdict \\
\midrule
Courtroom-month & 0.638 & 50 & $4.1\times10^{-30}$ & no clean lottery \\
Courtroom-week & 0.613 & 14 & $2.4\times10^{-16}$ & no clean lottery \\
Courtroom-day & 0.476 & 4 & $2.9\times10^{-11}$ & no clean lottery \\
Courtroom-day-hour & 0.405 & 3 & $3.3\times10^{-10}$ & no clean lottery \\
Courtroom-day by representation & 0.296 & 3 & 0.041 & limited diagnostic support \\
\bottomrule
\end{tabular}
\end{table}

\begin{figure}[H]
    \centering
    \includegraphics[width=0.78\linewidth]{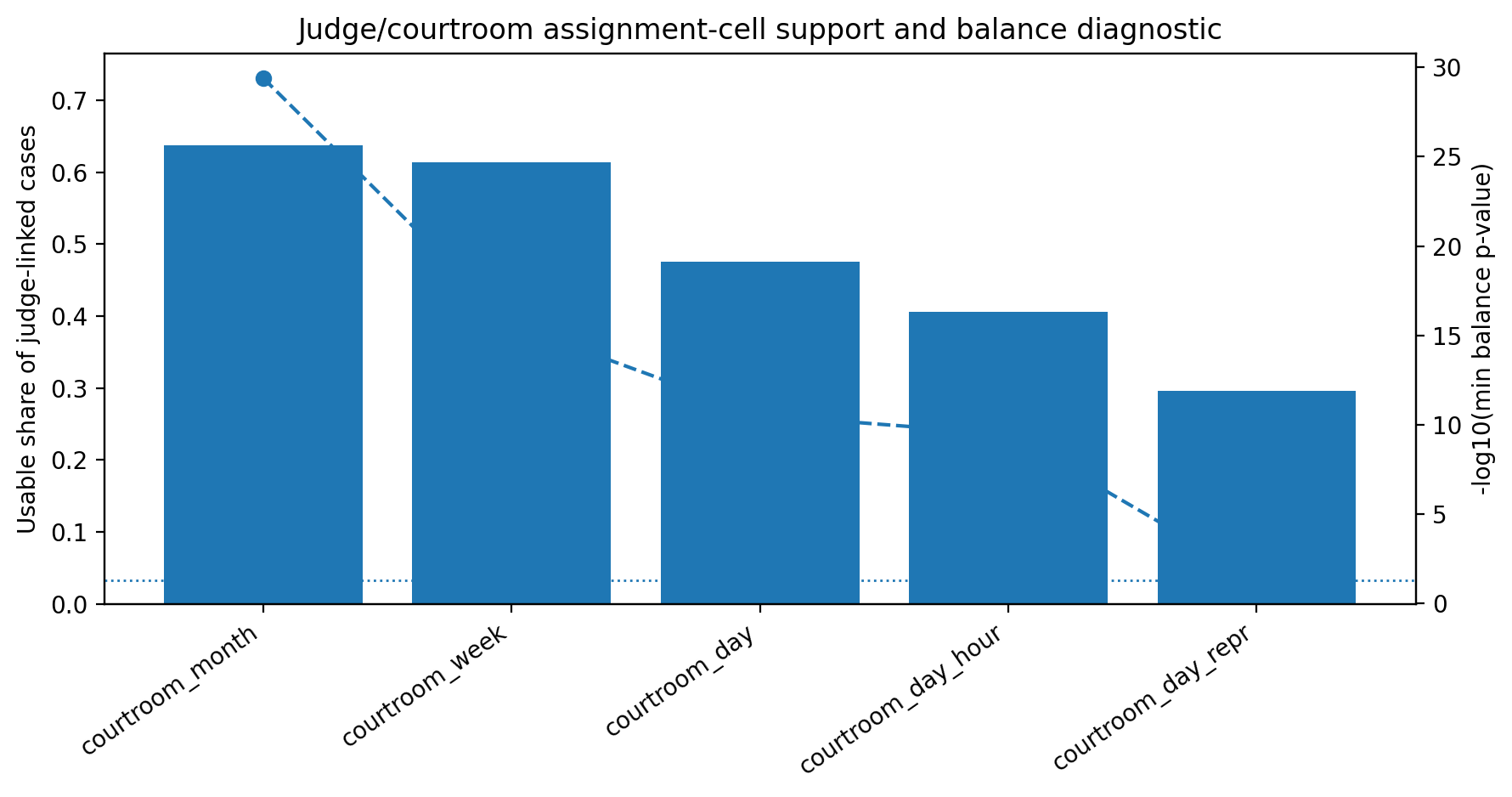}
    \caption{Judge/courtroom assignment-cell support and balance diagnostic. Broad cells retain more cases but fail balance; narrower cells improve comparability only with limited support.}
    \label{fig:judge_lottery_appendix}
\end{figure}

\subsection{Judge--triad support}\label{app:D2}

The judge $\times$ triad design is not estimable. The appendix support table finds 1{,}523 judge $\times$ triad groups, a maximum group size of four, and zero groups with at least five cases. This lack of support holds for the outcome-specific samples as well: no outcome has any judge $\times$ triad group with at least five observations. The correct interpretation is lack of estimability, not evidence of no effect.

\begin{table}[H]
\centering
\caption{Judge $\times$ triad support diagnostic.}
\label{tab:judge_triad_support_appendix}
\small
\begin{tabular}{p{0.36\linewidth}p{0.18\linewidth}p{0.18\linewidth}p{0.18\linewidth}}
\toprule
Unit or interaction & Groups & Maximum group size & Groups with $n\ge5$ \\
\midrule
Judge & 69 & 20{,}769 & 65 \\
Triad & 16{,}797 & 1{,}047 & 619 \\
Plaintiff--attorney & 8{,}182 & 2{,}172 & 822 \\
Judge $\times$ plaintiff--attorney & 1{,}387 & 20 & 6 \\
Judge $\times$ triad & 1{,}523 & 4 & 0 \\
\bottomrule
\end{tabular}
\end{table}

\begin{figure}[H]
    \centering
    \includegraphics[width=0.78\linewidth]{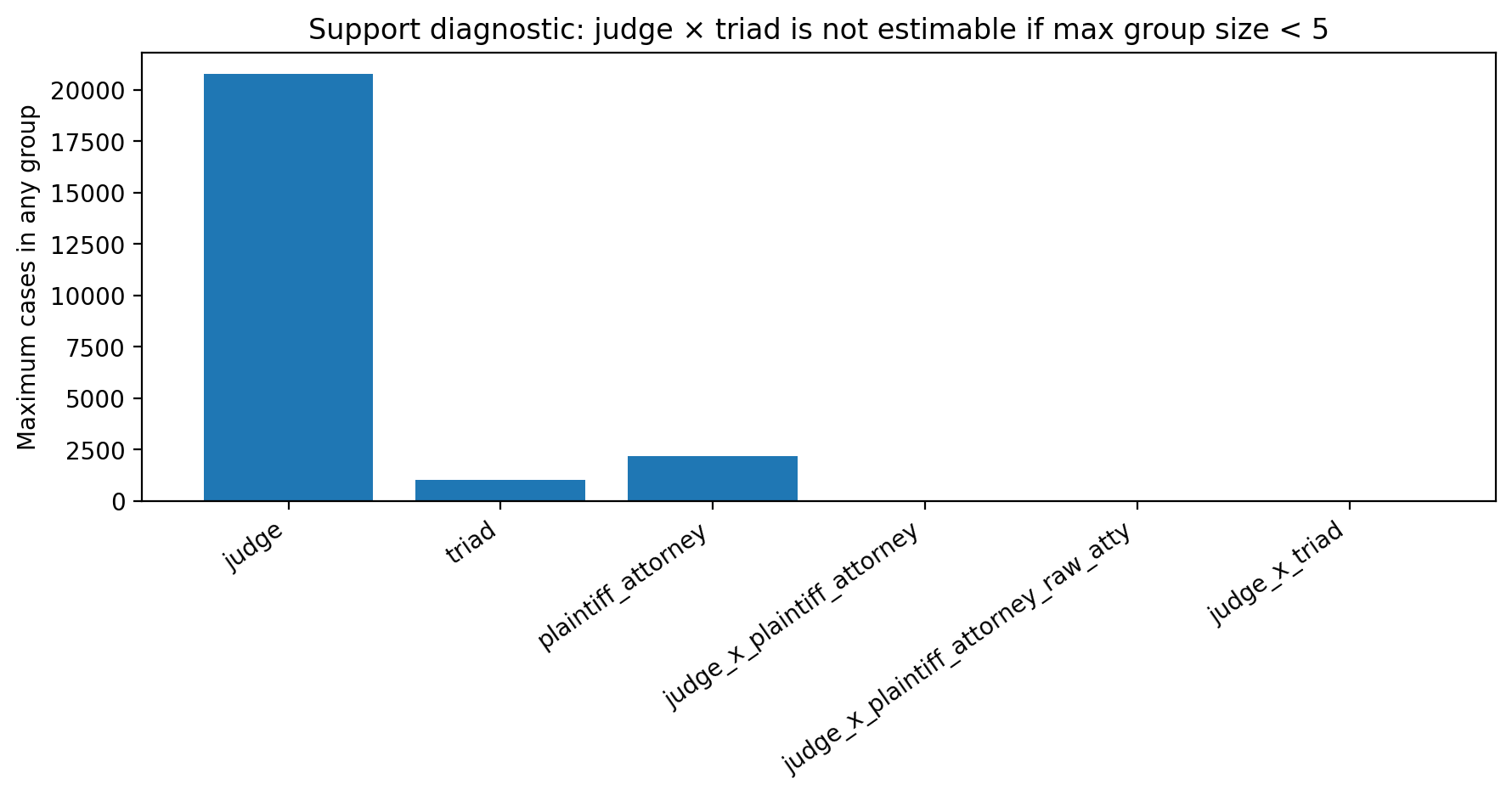}
    \caption{Support diagnostic for judge interactions. Judge $\times$ triad cells never reach five observations, so stable interaction effects cannot be estimated.}
    \label{fig:judge_triad_support_appendix}
\end{figure}

\subsection{Exploratory repeat-triad recurrence CSDID diagnostic}\label{app:D3}

The supplemental CSDID recurrence analysis treats the second observed filing of the same plaintiff--attorney--property triad as the event. The design is not used to claim randomized triad assignment. It is reported to test whether recurrent team--property units have an identifiable event-time pattern after the recurrence threshold is crossed. The only count outcome that is both statistically significant in the post-event average and not rejected by the pretrend screen is served-writ cases. For served writs, the recurrence-quarter ATT is \(+0.165\) (\(p<0.001\)), and the average ATT over quarters 0--8 is \(+0.0098\) (\(p=0.046\)). Other count outcomes are not interpreted as CSDID evidence because they fail the pretrend screen or do not produce a statistically reportable post-window estimate.

\begin{table}[H]
\centering
\caption{Repeat-triad recurrence CSDID diagnostic: statistically reportable outcome.}
\label{tab:triad_recurrence_csdid}
\small
\begin{tabular}{p{0.30\linewidth}p{0.20\linewidth}p{0.20\linewidth}p{0.20\linewidth}}
\toprule
Outcome & Event-quarter ATT & Post 0--8 ATT & Pretrend screen \\
\midrule
Served-writ cases & \(+0.165\), \(p<0.001\) & \(+0.0098\), \(p=0.046\) & not rejected \\
\bottomrule
\end{tabular}
\end{table}

\begin{figure}[H]
    \centering
    \includegraphics[width=0.70\linewidth]{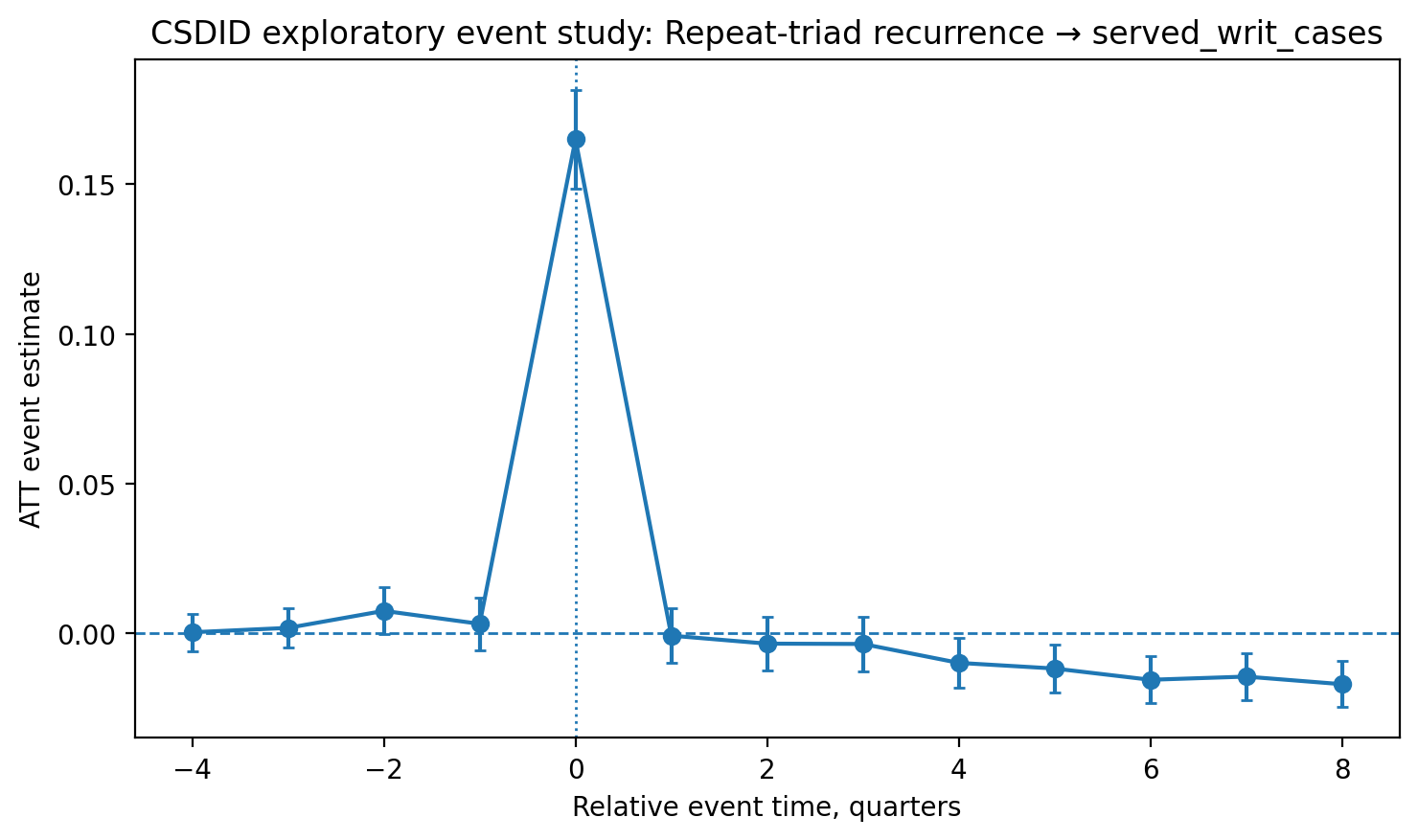}
    \caption{Repeat-triad recurrence CSDID diagnostic for served-writ cases. The figure is reported as a supplemental event-time diagnostic, not as evidence of randomized triad assignment.}
    \label{fig:triad_recurrence_csdid}
\end{figure}

\subsection{Triad-definition disclosure}\label{app:D4}

Section~4.4 uses the canonical triad construction. A separate integration check, designed for judge/courtroom diagnostics, produces a narrower repeat-triad universe of 2{,}712 groups versus the 52{,}443 in the canonical typology. The narrower construction is useful for diagnostic support checks and is not used as a replacement for the substantive triad typology.

\begin{table}[H]
\centering
\caption{Triad-definition disclosure.}
\label{tab:triad_definition_disclosure}
\small
\begin{tabular}{p{0.46\linewidth}p{0.20\linewidth}p{0.20\linewidth}p{0.10\linewidth}}
\toprule
Metric & Diagnostic construction & Canonical paper construction & Use \\
\midrule
Repeat-triad groups & 2{,}712 & 52{,}443 & diagnostic only \\
Repeat-triad cases & 17{,}914 & not targeted & diagnostic only \\
Section~4.4 triad typology & not replaced & retained & main text \\
\bottomrule
\end{tabular}
\end{table}

\end{document}